\documentclass[runningheads]{llncs}
\usepackage[T1]{fontenc}
\usepackage{amsmath}
\usepackage{graphicx}
\usepackage{xcolor}
\usepackage{graphicx}
\usepackage{soul}
\usepackage{tabularx}
\usepackage{caption}
\usepackage{subcaption}
\usepackage{float}
\usepackage{enumitem}
\usepackage{mwe} % to get dummy images

\begin{document}
\title{Trackerless freehand ultrasound with sequence modelling and auxiliary transformation over past and future frames}

\author{Qi Li\inst{1} \and
Ziyi Shen\inst{1} \and
Qian Li\inst{1,2}\and
Dean C Barratt\inst{1}\and
Thomas Dowrick\inst{1}\and
Matthew J Clarkson\inst{1}\and
Tom Vercauteren\inst{3}\and
Yipeng Hu\inst{1}
}
\authorrunning{Q. Li et al.}

\institute{Centre for Medical Image Computing, Wellcome/EPSRC Centre for Interventional and Surgical Sciences, Department of Medical Physics
and Biomedical Engineering, University College London, London, U.K. \and
State Key Laboratory of Robotics and System, Harbin Institute of Technology, Harbin, China \and
School of Biomedical Engineering \& Imaging Sciences, King’s College London, London, U.K.}
\maketitle              % typeset the header of the contribution
\begin{abstract}
Three-dimensional (3D) freehand ultrasound (US) reconstruction without a tracker can be advantageous over its two-dimensional or tracked counterparts in many clinical applications. In this paper, we propose to estimate 3D spatial transformation between US frames from both past and future 2D images, using feed-forward and recurrent neural networks (RNNs). With the temporally available frames, a further multi-task learning algorithm is proposed to utilise a large number of auxiliary transformation-predicting tasks between them. Using more than 40,000 US frames acquired from 228 scans on 38 forearms of 19 volunteers in a volunteer study, the hold-out test performance is quantified by frame prediction accuracy, volume reconstruction overlap, accumulated tracking error and final drift, based on ground-truth from an optical tracker. 
The results show the importance of modelling the temporal-spatially correlated input frames as well as output transformations, with further improvement owing to additional past and/or future frames.
The best performing model was associated with predicting transformation between moderately-spaced frames, with an interval of less than ten frames at 20 frames per second (fps).
Little benefit was observed by adding frames more than one second away from the predicted transformation, with or without LSTM-based RNNs. 
Interestingly, with the proposed approach, explicit within-sequence loss that encourages consistency in composing transformations or minimises accumulated error may no longer be required.
The implementation code and volunteer data\footnote[1]{\url{https://github.com/ucl-candi/freehand}} will be made publicly available ensuring reproducibility and further research.
\end{abstract}
\section{Introduction}
\label{sec:intro}

Reconstructing freehand ultrasound in 3D provides spatial information between acquired 2D frames, potentially for a wide range of clinical applications, and has indeed been adopted in areas including surgical and interventional guidance. In these applications, 3D reconstruction of the anatomy and pathology is essential for tasks such as registration to pre-operative imaging \cite{hu2016development} and quantifying 3D tissue motion \cite{jiang2021motion}.
It provides a low-cost, accessible alternative with larger and more flexible fields-of-view to the other 3D US imaging techniques such as 2D array transducer \cite{mozaffari2017freehand} and motorised probe.
Spatial tracking, electromagnetic or optical, is currently considered most robust approaches for 3D freehand US, but poses practical challenges for clinical adoption, due to the additional requirement such as extra equipment, line-of-sight or interference mitigation.
Therefore, tracker-free or image-based methods have generated long-lasting research interest, from previous work in exploiting physics-based speckle correlation models between image frames \cite{chen1997determination,hassenpflug2005speckle} to, more recently, machine learning-based methods \cite{prevost2017deep,ning2022spatial}, driven by supervising data often from spatial trackers for training.

Prevost \emph{et. al}. \cite{prevost20183d} proposed a convolutional neural network (CNN) to reconstruct 3D volume by estimating the transformation between two adjacent 2D images. FlowNet and densely connected networks were used in \cite{mikaeili2022trajectory,xie2021image}. In \cite{miura2020localizing}, ResNet and FlowNetS were integrated for a better localization and optical flow estimation, and consistency loss derived from stereo vision was added. Forward consistency loss was then proposed in \cite{miura2021probe}. In \cite{miura2021pose}, RNN was used to estimate both relative and absolute probe poses. In \cite{guo2020sensorless}, 3D CNN and Pearson correlation coefficient based case-wise correlation loss was proposed to enable more smooth trajectories.
A novel online learning framework with self-supervised learning method and adversarial training was proposed in \cite{luo2021self}.
The authors then integrated the IMU information both in training and inference time to extract velocity information and reduce drift error \cite{luo2022deep}.
Recently, \cite{ning2022spatial} used ResNet and transformer to extract local and global features of US sequence.   

This work builds on the previous effort in this challenging application and formulates the freehand US transformation estimation problem as a multi-task learning problem, not only focusing on the one transformation between a pair of images (main task) but a set of transformations (auxiliary tasks) between frames of the input image sequence. We show that this formulation is effective to capture strong correlation among input frames and that among output between-frame transformations.
It is a generalised algorithm that 
1) includes future frames in addition to past frames and their potential correlation;
and 2) predicts correlated neighbouring transformations in addition to the main task and takes advantage of cyclic and accumulative consistency between them. 

% The network encodes a number of frames as input, which has rich spatial information compared with previously proposed two adjacent frames. The image sequence encoding strategy allows the predicted transformation not limited to transformation of two adjacent frames. Instead, the output of network can be the transformation between any two frames in a input sequence. In addition, the effects of interval and position of the predicted transformation can be studied in this pipeline, which shows the influence of past and future frames to the current transformation prediction accuracy. 

Our contributions include: 1) a new design of freehand US sequence encoding in a novel multi-transformation learning algorithm; 2) extensive experimental results to quantify the benefits from the proposed methodological components; and 3) code and volunteer data for public access.
 
\section{Method}
\label{sec:method}
For an US scan consisting of a set of 2D image frames $\mathcal{S}$, image sequences with a length of $M$ can be sampled $S=\{I_m\}, m=1,2,...,M$, where $S 	\subseteq {\mathcal{S}}$ and $m$ denotes consecutively increasing time-steps at which the frames are acquired. For a given sequence, a spatial transformation $T_{j\leftarrow i}, 1 \leq i<j \leq M$ denotes the relative translation and rotation between the $i^{th}$ and $j^{th}$ frames.
This section describes our proposed method to predict the spatial transformation  $T_{j^*\leftarrow i^*}$ between a pair of frames $(i^*,j^*)$, with a $j^*-i^*$ interval, $i^*-1$ past frames and $M-j^*$ future frames. 

After models are trained by sequences randomly sampled from training US scans, a test scan can then be reconstructed by consecutively predicting multiple sequences with a predefined $M$, such that the $(j^*)^{th}$ frame from the previous sequence is the $(i^*)^{th}$ frame in the subsequent sequence, for the entire scan with variable length. All frames after the initial $j^*$ can be spatially localised with respect to their varying starting reference frame. Different values of $M$ are tested to include potentially useful long-term dependency.

%For a US scan containing $M$ 2D frames, $US = [I_i|i=0,1,2,...,M-1]$, where $I_i$ denotes $i^{th}$ frame in a US scan, the 3D US volume and the trajectory of the US probe can be reconstructed if the position of each frame is known, which can be computed using rigid transformations encoded as homogeneous matrices $T$ or parameter vectors $\hat{\theta}$ $\in$ $\R$. Particularly, as there is no absolute coordinate system, relative transformation estimation between two adjacent frames is equivalent to this problem.  

%An overview of the proposed method is illustrated in Fig.1. The network can be any convolutional neural network, e.g., efficientnet\_b1 network, which takes an image sequence as input and output corresponding transformations in a multi-task learning pipeline. The input image sequence is a subset of the entire scan $S = [I_i|i=m,m+1,m+2,...,n]$, where $m$ and $n$ are the start and end frames of an image sequence. The 3D volume can be reconstructed by using any transformation of the network prediction. 

\subsection{Input sequence encoding}
\label{ssec:Input sequence encoding}
A recurrent neural network $f_{rec}$ with parameters $\theta$ takes the image frames in sequence to predict $T_{j^*\leftarrow i^*}$:
\begin{align} \label{eq:rnn}
    T_{j^*\leftarrow i^*} &=  f_{rec}(I_m,h^{(m-1)};\theta), for~ m=M \\
    h^{(m)} &= f_{rec}(I_m,h^{(m-1)};\theta), \forall~m\leq{M-1}
\end{align}
where $h^{(m)}$ is the internal hidden state at time-step $m$ and the transformation $T_{j^*\leftarrow i^*}$ is predicted at the end of each sequence. Here, the future frames are used if $j^*<M$, leading to a time-delayed transformation prediction. Feed-forward CNN $f_{fwd}$ are also tested to model the same image sequence without considering the sequential steps explicitly:
\begin{align} \label{eq:cnn}
T_{j^*\leftarrow i^*} &=  f_{fwd}(S;\theta).
\end{align}

\begin{figure}[ht]

  \includegraphics[width=0.8\linewidth]{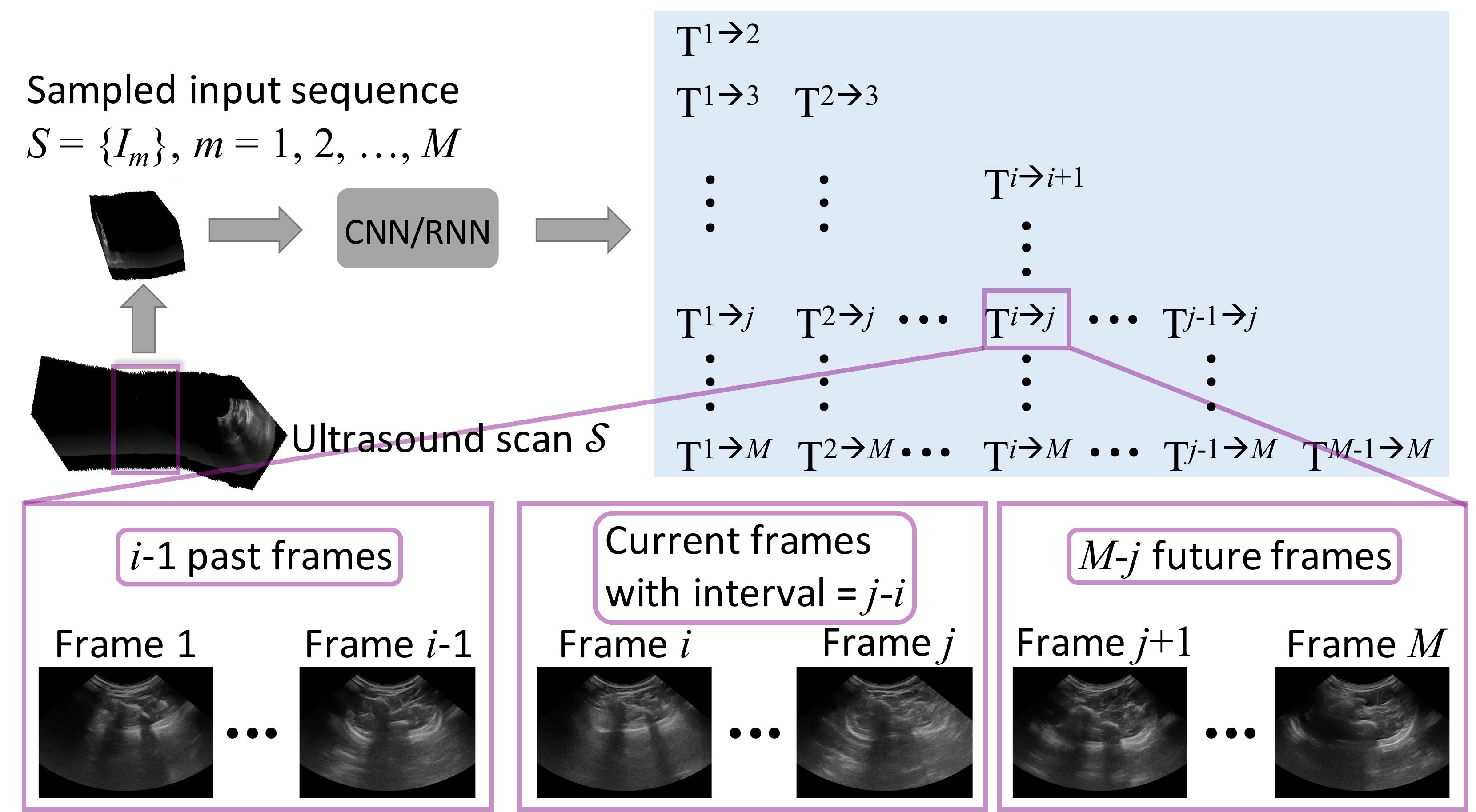}
  \centering
  \caption{Illustration of input sequence encoding and multi-task output in the proposed method.}
  \label{fig:input_output}
\end{figure}

Given a predefined pair indices $(i^*,j^*)$ and the sequence length $M$, this formulation includes permutations of available neighbouring frames $I_{m\in ([1,i^*-1] \cup [j^*+1,M])}$ and their relative positions, as shown in the example in Fig.~\ref{fig:input_output}.

%\vspace{-10}

%\subsection{Loss functions}
%\label{ssec:Loss function}  
%
In this work, we propose to consider $(i^*,j^*)$ as hyperparameters, tuned on validation set. This together with $M$ is equivalent to a flexible, generalised frame-encoding that provides a conditioning context for spatial transformation prediction. 
For example, smaller $i^*$ and $M-j^*$ indicate prediction using a shorter history and fewer future frames, respectively; and a single-pair input is represented by $M=2$, found in several previous studies. The benefits of an effective and efficient context-enabled encoding have been studied in related areas such as n-gram encoding \cite{takase2019character}. 

Efficient frame encoding is particularly important in this application due to memory required, for both feed-forward and unfolded recurrent networks, with the high-dimensional image input and potentially long US sequence, and, as shown in this study, may warrant much shorter sequences required for tracking and scan reconstruction. Furthermore, this simultaneously enables a practically data structure for the multi-task learning described in Sec.~\ref{ssec:Multi-task learning}.

%Image sequence, a volume of frames in a scan sequence, has richer spatial information than two adjacent 2D images as it reflects displacement through time. Let $T^{i-1\to i}$, the 4$\times$4 transformation matrix which consists of a 3$\times$3 rotation matrix and a 3$\times$1 translation vector, be the transformation from frame $i-1$ to frame $i$. Previous works usually only use frames $i-1$ and $i$ to estimate transformation $T^{i-1\to i}$, lacking global spatial information of the scan sequence. In this work, for $T^{i-1\to i}$, not only current frames $i-1$ and $i$, but also the past frames, e.g., frames $i-2$, $i-3$,... and future frames, e.g., frames $i+1$, $i+2$,... are used to formulate a joint probability of the spatial transformation. Especially, the estimated transformation can not only be adjacent $T^{i-1\to i}$, but also can have discontinuous transformation $T_{i\to{j}}$, where $j-i>1$. Based on the n-gram theory, the probability of current transformation, instead of associating with all the frames in a scan sequence, can be approximated by using only their neighboring frames, which makes sequence encoding aligned with the whole pipeline. In experiments part, experiments are designed to detected the effects of the number of past frames and future frames. 

\subsection{Multi-task learning}
\label{ssec:Multi-task learning}

Whilst predicting $T_{j^*\leftarrow i^*}$ is regarded as the main task, both the recurrent and feed-forward networks can be adapted to predict other transformations $T_{j\leftarrow i}, i\neq{i^*} or~ j\neq{j^*}$. This work proposes to predict all these possible $C_M^2-1$ transformations as auxiliary tasks, also illustrated in Fig.~\ref{fig:input_output}. The differences between the predicted $\hat{T}_{j\leftarrow i}$ and ground-truth $T^{(gt)}_{j\leftarrow i}$ can be averaged as the overall loss for network training. When $C_M^2$ is very large, randomly selected $\tau$ samples of the auxiliary tasks may be used instead, $\tau \leq C_M^2-1$. 

The proposed multi-task learning for the ``neighbouring transformations'', albeit conceptually simple, not only exploits the shared representation from the auxiliary transformation prediction tasks, but also facilitates other losses based on these correlated transformations, such as the consistency loss and the alternative accumulated loss in Sec.~\ref{ssec:Loss function}. 
%As the network input is image sequence, the output of network can be transformations between any two frames in the input image sequence, and each prediction can be regarded as one task in the multi-task learning pipeline. The number of tasks can be $C_n^2$. Each transformation is determined by its position within the input sequence and the transformation interval which is the span of the investigated two frames. For input sequence $S$, $T_{i\to{j}}$, with interval $j-i$, is estimated by using $i$ past frames and $n-1-j$ future frames. Fig. 2 shows an example of the network input and output in one step. The input of network is 7 frames, i.e., frames 0, 1, 2,...6, and the network output is $C_7^2 = 21$ transformations between any two frames of input sequence, e.g., transformation from frame 3 to frame 5, $T^{3\to 5}$. For prediction $T^{3\to 5}$ with interval = 2, frames 0, 1, 2 are regarded as past frames, and frame 6 is the future frame.
\subsection{Loss functions}
\label{ssec:Loss function}  
To alleviate empirical tuning between rotational and translational contributions, this work adopts loss functions based on the distance between the prediction-transformed points $\hat{p}^{(j)}_n, n=1,...,N$ and the ground-truth-transformed points $p^{(j)}_n$, by the predicted $\hat{T}_{j\leftarrow i}$ and ground-truth $T^{(gt)}_{j\leftarrow i}$, respectively. In this work, $N=4$ corner points in $j^{th}$ image are used, in their homogeneous tracking \textit{tool space}. Therefore, the multi-task loss function is the average of the mean-square-errors (MSEs) over the $\tau+1$ tasks:
%Since three rotation and three translation in parameter vector $\hat{\theta}$ are not in the same numerical range, with magnitude $10^{-3}$ and $10^{-1}$ respectively, the transformed points distance can be used to measure the dissimilarity between the ground-truth and network prediction. The loss function is the mean squared error (MSE) of the coordinates of pixels in a frame between the ground-truth and prediction:        
\begin{equation} \label{eq:loss-multitask}
\mathcal{L}_{multi-task} = \frac{1}{N\cdot(\tau+1)} \sum^{\tau+1}\sum_{n=1}^{N}\emph{D}(p^{(j)}_{n},\hat{p}^{(j)}_{n})
\end{equation}
where $\emph{D}(\cdot)$ denotes MSE between $x$, $y$ and $z$ coordinates of the two point sets, $\hat{p}^{(j)}_{n} = \hat{T}_{j\leftarrow i} \cdot  T_{(calib)} \cdot p^{(i)}_{n}$,
and 
$p^{(j)}_{n} = T^{(gt)}_{j\leftarrow i} \cdot T_{(calib)} \cdot p^{(i)}_{n}$. $p^{(i)}_{n}$ represents the same points in the $i^{th}$ \textit{image space}, and $T_{(calib)}$ is a fixed transformation from image space to tool space, obtained through calibration. 

The ground-truth transformation is composed by two tool-to-world transformations,
$T^{(gt)}_{j\leftarrow i}=(T^{(gt)}_{world\leftarrow j})^{-1} \cdot T^{(gt)}_{world\leftarrow i}$, at the time-steps $i$ and $j$, obtained from the optical tracker, thus independent of the \textit{world (camera) space}. A further left-multiplication by ($T_{(calib)})^{-1}$ could compute distance defined in the image space, should it be preferred.

Among the predicted transformations, consistency may be enforced between a direct prediction $\hat{T}_{j\leftarrow i}$ and an indirect prediction $\hat{T}^{\oplus}_{j\leftarrow i} = \hat{T}_{j\leftarrow k} \cdot \hat{T}_{k\leftarrow i}$.
Given each time-step $k$ and additional transformations to and from $k$, a set of consistency losses on the transformed points can be defined for each task:
\begin{equation} \label{eq:loss-consistency}
\mathcal{L}_{consistency} = \frac{1}{N}\sum_{n=1}^{N}\emph{D}(\hat{p}^{(j)}_{n},\hat{p}^{(j)\oplus}_{n})
\end{equation}
where $\hat{p}^{(j)\oplus}_{n} = \hat{T}^{\oplus}_{j\leftarrow i} \cdot  T_{(calib)} \cdot p^{(i)}_{n}$. 
This loss function only promotes consistency and does not require ground-truth data. It should be used in conjunction with Eq.~\ref{eq:loss-multitask} (or Eq.~\ref{eq:loss-accumulated}) to avoid trivial solutions. 
Importantly, the consistency loss is a form of ``teacher forcing'' commonly adopted in training sequence models, which makes use of the ground-truth targets, rather than the previous predictions, to supervise the subsequent prediction during training. It has been proven advantageous in sequence-to-sequence models \cite{miura2021probe}.

Alternatively, minimising the difference between $\hat{p}^{(j)\oplus}_{n}$ and ground-truth $p^{(j)}_{n}$ forms an accumulated loss:
\begin{equation} \label{eq:loss-accumulated}
\mathcal{L}_{accumulated} = \frac{1}{N}\sum_{n=1}^{N}\emph{D}(p^{(j)}_{n},\hat{p}^{(j)\oplus}_{n})
\end{equation}

%One frame can be transformed to another frame by using various transformation combinations, and can be formulated by consistent loss and accumulated loss. For example, points in frame $c$ can be transformed from frame $a$ by transformation $T^{a\to b}$ and then $T^{a\to b}$, i.e., transformation $T^{b\to c}$\cdot$T^{a\to b}$, or $T^{a\to c}$ directly. (will update)   

Although not investigated in this work, finding the optimal relative weighting between these loss terms should further improve the proposed method and be of interest in future studies. 
The reported results in Sec.~\ref{sec:Experiments} used equal weighting to provide a reference performance.

\subsection{Evaluation metrics}
\label{ssec:Evaluation metrics}

For each sequence, the Euclidean distance between prediction and ground-truth on four corner points of consecutive frames is defined as \textit{frame prediction accuracy} ($\epsilon_{frame}$), which assesses model generalisation without scan reconstruction.

For each reconstructed scan, two reconstruction errors are reported: 1) an \textit{accumulated tracking error} ($\epsilon_{acc.}$) is the average Euclidean distance over all reconstructed image pixel locations; and 2) a \textit{volume reconstruction overlap} ($\epsilon_{dice}$) measure, Dice between the reconstructed volumes of prediction and ground-truth, where a reconstructed volume is approximated with hexahedrons formed by two adjacent frames. A \textit{final drift} ($\epsilon_{drift}$) is also reported as the Euclidean distance, averaged over the four corners, between the final predicted and ground-truth frames in each scan.

All results are reported on the hold-out test set, unseen to model training and development. These error metrics are designed for a range of freehand US applications that may have different clinical focuses \cite{mozaffari2017freehand}.

\iffalse

From the perspective of overall accuracy of the entire scan, different clinical scenarios has different expectation for 3D freehand US reconstruction. One the one hand, some applications expect precise trajectory of the entire scan, for example, the vessel length measurement and organ volume measurement application. In these applications, the average accumulated error can be used to measure the trajectory accuracy:

\begin{equation} \label{accumulated_error}
avg.\, acc.\, error = \frac{1}{M}\sum_{j=0}^{M-1}\frac{1}{N}\sum_{i=0}^{N-1}\|p_{i,j}^{(gt)}-\hat{p}_{i,j}\|_2^\frac{1}{2}
\end{equation}

On the other hand, in volume measurement application, 3D volume reconstruction accuracy becomes the principal metric, which can be computed by using volume overlap and average distance of all pixels between ground-truth volume and prediction volume:

\begin{equation} \label{volume_overlap}
vol.\, overlap = \emph{O}(V^{(gt)}-\hat{V})
\end{equation}
where $\emph{O}(\bullet)$ computes the volume overlap between ground-truth volume $V^{(gt)}$ and prediction volume $\hat{V}$.

\begin{equation} \label{average_all_pxiels_dists}
avg.\, pixel.\, dists = \sum_{j=0}^{M-1}\frac{1}{m}\sum_{i=0}^{m-1}\|p_{i,j}^{(gt)}-\hat{p}_{i,j}\|_2^\frac{1}{2}
\end{equation}
where $m$ is the number of pixels in a frame. 

\fi

\section{Experiments and results}
\label{sec:Experiments}
\subsection{Data acquisition}
\label{ssec:Data_acquisition}
Freehand US scans were acquired on both left and right forearms from 19 volunteers. On each forearm, the US probe was moved, for study purpose, in a straight line, a `C' shape and a `S' shape, in a distal-to-proximal direction. These three scans were repeated, with the curved-linear transducer held (thus the US planes) perpendicular of and parallel to the forearm. After manually cropping the initial and end stages when the probe was largely stationary, between 36 and 430 frames with a size of 480×640 pixels, equivalent to a probe travel distances between 100 and 200 mm, were included. One scan with less than 50 frames was discarded for its uncertain quality. The data was split into train, validation and test sets by a ratio of 3:1:1, without the same forearm in different sets. All US scans were acquired on Ultrasonix machine (BK, Europe) with a curve-linear probe (4DC7-3/40), tracked by an NDI Polaris Vicra (Northern Digital Inc., Canada). B-mode images with median level of speckle reduction were recorded at 20 fps. Spatial (image-to-tool) and temporal differences were calibrated using a pinhead-based method \cite{hu2017freehand} and the Plus Toolkit~\cite{lasso2014plus}, respectively.

\subsection{Network development and implementation}
\label{ssec:model_development}
This work aims to provide an established network performance, without focusing on further architecture optimisation. The EfficientNet (b1) \cite{tan2019efficientnet} was adapted as the feed-forward CNN, with a no-activation output layer to predict $(\tau+1) \times 6$ dimensional vectors representing the multi-task predictions. The same EfficientNet-based feature encoder followed by a long short-term memory (LSTM) module \cite{hochreiter1997long}, with a 1024-dimensional hidden feature vector, was used as the recurrent network. A baseline CNN was also trained, with two adjacent frames as input and output transformation between them. 
A minibatch size of 32 and the Adam optimizer were used to train each model for 50,000 epochs. The best model with the minimum frame prediction accuracy on the validation set was selected, and then report the results on the test set. 
In addition to the network and training options described above and those in Sec.~\ref{sec:method}, other hyperparameter values including a learning rate of $10^{-4}$, tested among $\{10^{-3}, 10^{-4}, 10^{-5}\}$, and a sequence length of 20, tested among $\{10, 20, 30, 40, 49\}$, were selected with $\tau=79$ based on the validation set performance.

%The LSTM-based RNN consists of encoder part and decoder part. The encoder part aims to extract the feature vector for each frame in input sequence, based on the same network architecture in the baseline. The decoder part takes the extracted feature of one frame as input at each time step, and output all the transformations in the final time step.  

%A number of experiments were designed to analysis the  hyper-parameters included in the proposed framework, such as the number of frames in the input sequence, the difference between using CNN and RNN, and the effects of past frames, future frames, and interval. Input sequence length of 10, 20, 30, 40, and 49 were selected for comparison. The output tasks are sampled from the original $C_n^2$ tasks with the trade off between training time and task diversity, containing transformations with various intervals, number of past and future frames. Specifically, when computing the number of past and future frames, the current transformation is regarded as a whole. For example, for transformation from frame 5 to 8 with an input sequence length 10, the interval, number of past and future frames are 3, 5, and 1 respectively.      

%The following experiment results are based on models with sequence length 30 and learning rate 1e-4 as this setting gives a better performance than the other settings.

 %As not all the tasks in this multi-task learning pipeline give a good performance, which is illustrated in the above section, a number of transformation tasks with relatively good performance are selected for comparison in this section.

\subsection{Comparison to the baseline and ablation study}
\label{ssec:Ablation_study}
On the hold-out test set, ablation studies quantify the impact on the performance due to 1) the addition of auxiliary tasks, 2) the number of past frames and future frames included in input sequence; 3) the frame interval $j-i$ between which the transformation is predicted; and 4) the choice between feed-forward CNNs and LSTM-based RNNs.

\begin{figure}[h!]
  \includegraphics[width=0.9\columnwidth]{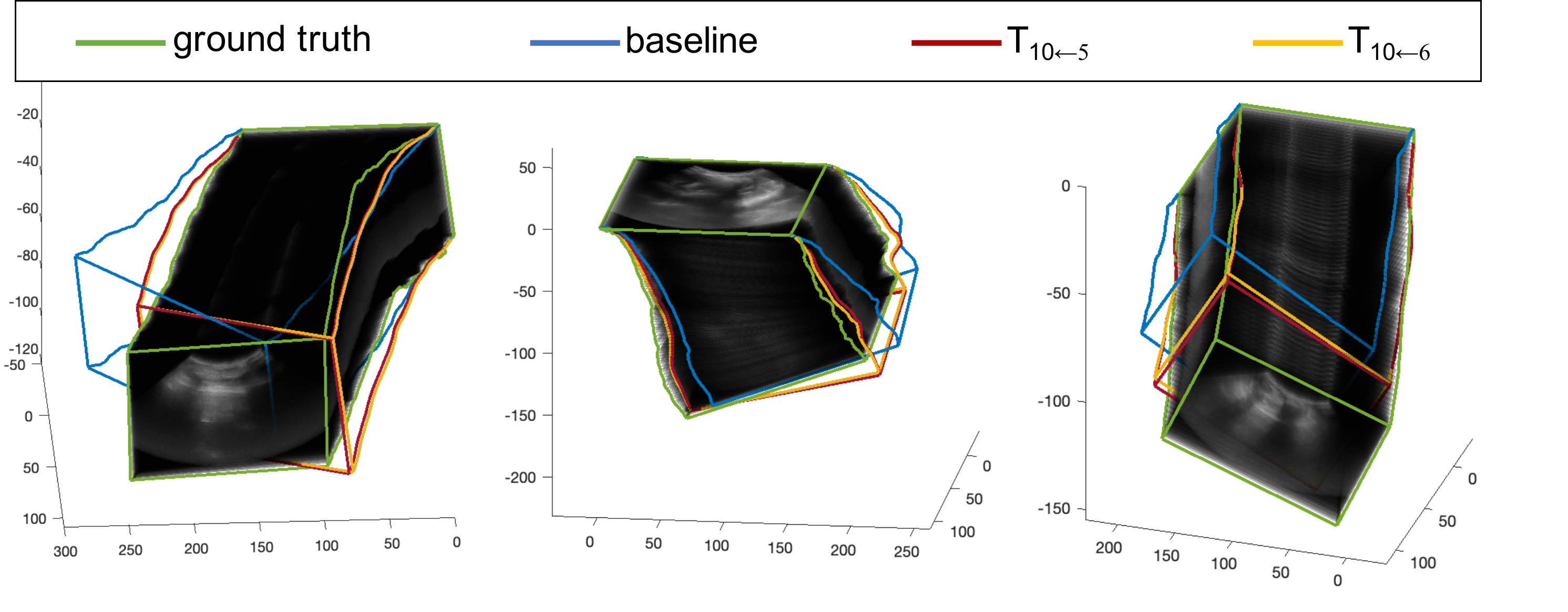}
  \centering
  
  \caption{Reconstruction results of baseline and proposed method (using $T_{10\leftarrow 5}$ and $T_{10\leftarrow 6}$).}
  \label{fig:reconstruction_results}
\end{figure}

As shown in Table.~\ref{table:Evaluation_baseline} and Fig.~\ref{fig:reconstruction_results}, both $\epsilon_{frame}$ and $\epsilon_{acc.}$ were improved after adding auxiliary tasks, regardless their permutations, compared with the baseline (p$\leq$0.001 for $\epsilon_{frame}$, $\epsilon_{acc.}$, $\epsilon_{drift}$ and p$\leq$0.033 for $\epsilon_{dice}$, paired t-tests at $\alpha$=0.05), where $\epsilon_{acc.}$ increases with time, whilst $\epsilon_{frame}$ is relatively stable between sequence locations in the scan. $\epsilon_{dice}$ was computed on the perpendicular scans as an example.     %Note: 1) as the input sequence in the proposed method is selected as 20, which means that when reconstructing a scan, the last input sequence whose length is less than 20 will not be reconstructed; 2) $T_{10\leftarrow 5}$ in Table.~\ref{table:Evaluation_baseline} means using the transformation from 5$^{th}$ frame to 10$^{th}$ frame to reconstruct the scan. 
%2) as the computing time for overlap is very long, in this paper, we compute this metric by using the first 100 frames in a scan, which is enough for comparison use.
\begin{table*}[t!]
\tiny

\caption{Reconstruction performance of baseline and proposed method.}
% title of Table
\centering
%\arraybackslash
% used for centering table
%\resizebox{\columnwidth}{%
\begin{tabular}[b]{c c c c c c c c}
% centered columns (4 columns)
%{c c c c c c c c c c}
\hline
Evaluation metrics(mm) & Baseline & $T_{6\leftarrow 1}$ & $T_{10\leftarrow 5}$ &  $T_{9\leftarrow 6}$ &  $T_{10\leftarrow 6}$& $T_{9\leftarrow 6}^{accumulated}$& $T_{9\leftarrow 6}^{consistency}$\\ %[8pt]
% inserts table
%heading
\hline
% inserts single horizontal line
$\epsilon_{frame}$-cnn & $0.63\pm0.54$ & $0.55\pm0.57$  & $0.53\pm0.56$ & $0.57\pm0.57$ & $0.55\pm0.56$&$0.58\pm0.60$  & $0.58\pm0.59$\\
$\epsilon_{frame}$-LSTM & $0.66\pm0.46$ & $0.53\pm0.42$ & $0.50\pm0.41$ & $0.53\pm0.43$ & $0.51\pm0.41$ &  $0.54\pm0.44$& $0.56\pm0.48$ \\
% inserting body of the table
%$\epsilon_{acc.}$ & 24.42\pm17.17|27.91\pm15.37 & 20.52\pm11.87|20.15\pm9.41 & 20.59\pm11.53|19.98\pm10.97  & 20.32\pm12.07|20.35\pm11.42 & 20.39\pm12.29|20.05\pm11.03& XX & XX\\

$\epsilon_{acc.}$-cnn & $24.42\pm17.17$  & $19.05\pm13.64$ & $19.09\pm14.60$ & $19.03\pm13.68$ & $19.15\pm14.33$ & $20.94\pm14.58$ & $20.98\pm15.18$ \\ %[1ex]
$\epsilon_{acc.}$-LSTM & $27.91\pm15.39$ & $19.18\pm10.19$ & $18.13\pm9.49$ & $18.21\pm9.18$ & $18.56\pm9.65$ & $20.35\pm9.82$ & $20.52\pm13.54$\\

$\epsilon_{dice}$-cnn & $0.72\pm0.22$&$0.80\pm0.11$ & $0.81\pm0.11$ & $0.79\pm0.12$ & $0.80\pm0.11$ & $0.76\pm0.20$& $0.75\pm0.22$  \\ %[1ex]
$\epsilon_{dice}$-LSTM & $0.68\pm0.20$ & $0.76\pm0.12$ & $0.78\pm0.12$ & $0.78\pm0.11$ & $0.78\pm0.11$& $0.65\pm0.48$ &$0.76\pm0.15$ \\
$\epsilon_{drift}$-cnn& $46.01\pm33.34$ & $37.96\pm27.98$ & $36.82\pm28.01$ & $37.19\pm27.37$ & $36.93\pm27.63$ & $42.33\pm27.48$ & $40.54\pm30.64$\\
$\epsilon_{drift}$-LSTM& $51.43\pm30.30$ & $40.56\pm24.29$ & $36.48\pm20.77$ & $37.26\pm20.73$ & $37.36\pm20.75$ & $40.33\pm22.09$ & $39.50\pm27.03$\\[0.5ex]
% [1ex] adds vertical space
\hline
%inserts single line
\end{tabular}
%}
\label{table:Evaluation_baseline}
% is used to refer this table in the text
\end{table*}

Fig.~\ref{fig:interval_pf_ff} plots the performance in $\epsilon_{acc.}$ versus variable intervals, number of past and future frames, respectively. It shows that a relatively short interval, for both CNN and LSTM, between 3 and 9, resulted lower errors (e.g. unpaired p=0.010, LSTM at interval=9 vs. baseline). The use of past and future frames was clearly beneficial, compared with those without, i.e., $x=0$ in Fig.~\ref{fig:interval_pf_ff} b and c. However, an interesting observation is that, performance improved when $<$5 past frames was added, whilst additional 9-11 future frames values offered lowest $\epsilon_{acc.}$.   
%However, an interesting observation is that, performance improvement was only observed when 1-5 past frames being added, whilst the benefit of additional future frames was more monotonic, with optimal values between 9 and 11, lasting to what was considered allowed system delay, here $\sim$30 frames for 1-2 second. 
The need for longer-term dependency was unsubstantiated, for example no significant improvement was found by increasing the sequence length beyond 20 during model development. However, the RNNs yielded consistent lower prediction variance, as shown in Fig.~\ref{fig:interval_pf_ff}, which may indicate a superior within-sequence modelling. $\epsilon_{dice}$ and $\epsilon_{drift}$ showed consistent conclusions to those based on $\epsilon_{acc.}$, therefore omitted for brevity in the plots. 

\begin{figure*}[h!]
\centering
  \includegraphics[width=\linewidth]{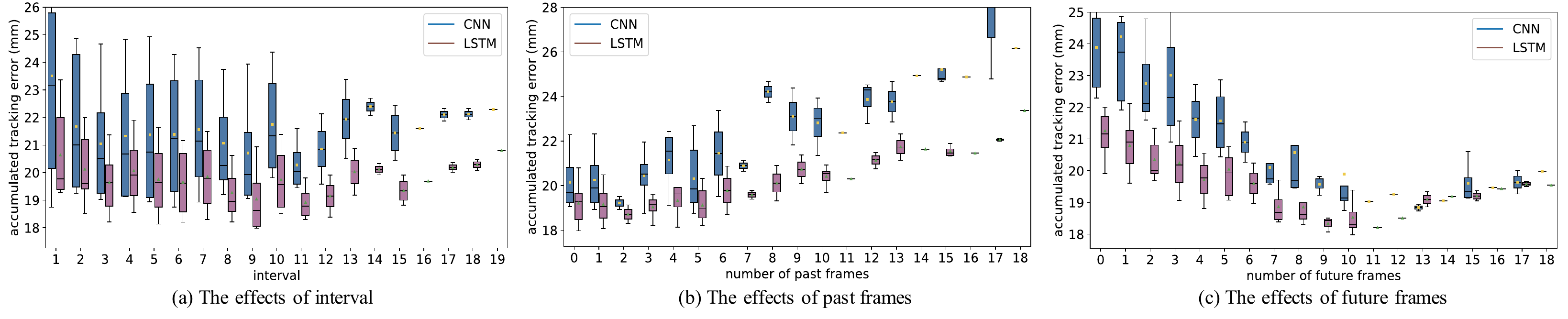}
  
  \caption{The performance of accumulated tracking error with various intervals, number of past and future frames.}
  \label{fig:interval_pf_ff}
\end{figure*}

\iffalse
\begin{figure}[!ht]
\centering
\begin{subfigure}{\columnwidth}%{\textwidth}
  \centering
  \includegraphics[width=\linewidth]{acc_err_interval.pdf}
  
  \caption{The effects of interval}
  
  \label{fig:interval}
\end{subfigure}%
\hfill
\begin{subfigure}{\columnwidth}%{\textwidth}
  \centering
  \includegraphics[width=\linewidth]{acc_err_pf.pdf}
  
  \caption{The effects of past frame}
  
  \label{fig:pf}
\end{subfigure}
\begin{subfigure}{\columnwidth}%{\textwidth}
  \centering
  \includegraphics[width=\linewidth]{acc_err_ff.pdf}
 
  \caption{The effects of future frame}
  
  \label{fig:ff}
\end{subfigure}
\caption{The performance of frame prediction accuracy with various intervals, number of past and future frames.}
\label{fig:interval_pf_ff}
\end{figure}
\fi

Fig.~\ref{fig:frame_prediction_accuracy} plots mean and variance of $\epsilon_{frame}$ and $\epsilon_{acc.}$, over all scans in the test set, between baseline and the proposed CNN-based multi-task model. As an example in predicting $T_{10 \leftarrow 6}$, the improvement from the multi-task learning seems increased as the sequences accumulate. 
%Note that the high variance both in baseline and proposed method may be due to the various scan length in the dataset.
%As most of scan length is no greater than 300, Fig.~\ref{fig:frame_prediction_accuracy} only shows the first 300 frames, illustrating that the proposed multi-task learning framework do have optimal task with the help of multi-task. 
\iffalse
\begin{figure}
\begin{minipage}[b]{1.0\linewidth}
  \includegraphics[width=\linewidth]{frame_prediction_accuracy.pdf}
  
  \caption{Comparison of $\epsilon_{frame}$ and accumulated $\epsilon_{frame}$ between baseline and proposed method.}
  \label{fig:frame_prediction_accuracy}
 \end{minipage}

\end{figure}
\fi
\begin{figure}[!ht]
%\centering
\begin{subfigure}{.45\columnwidth}%{\textwidth}
  \centering
  \includegraphics[width=1.09\textwidth]{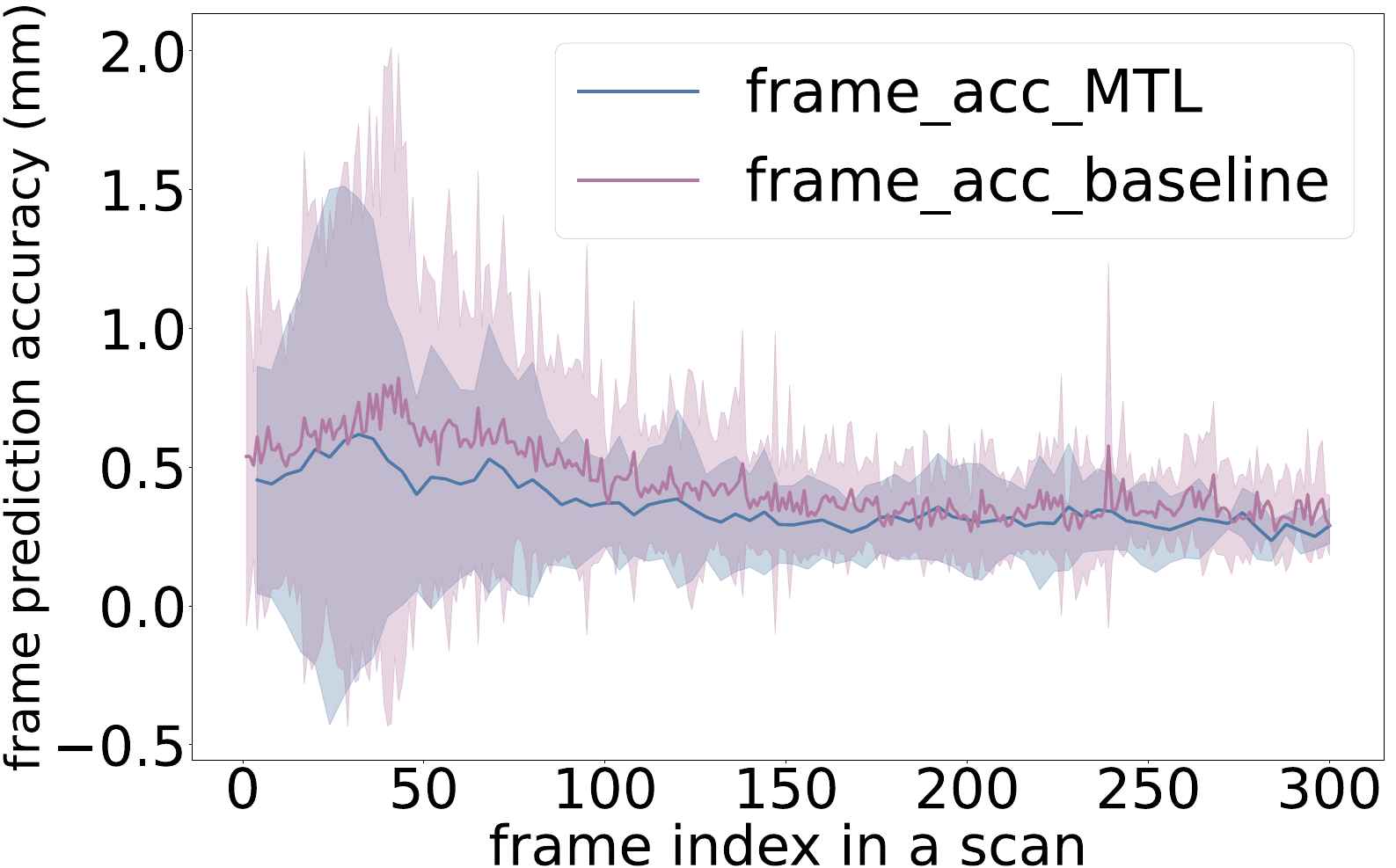}
 
  \caption{$\epsilon_{frame}$}
  
  \label{fig:frame_err}
\end{subfigure}%
\hfill
\begin{subfigure}{.47\columnwidth}%{\textwidth} linewidth
  \centering
  \includegraphics[width=\textwidth]{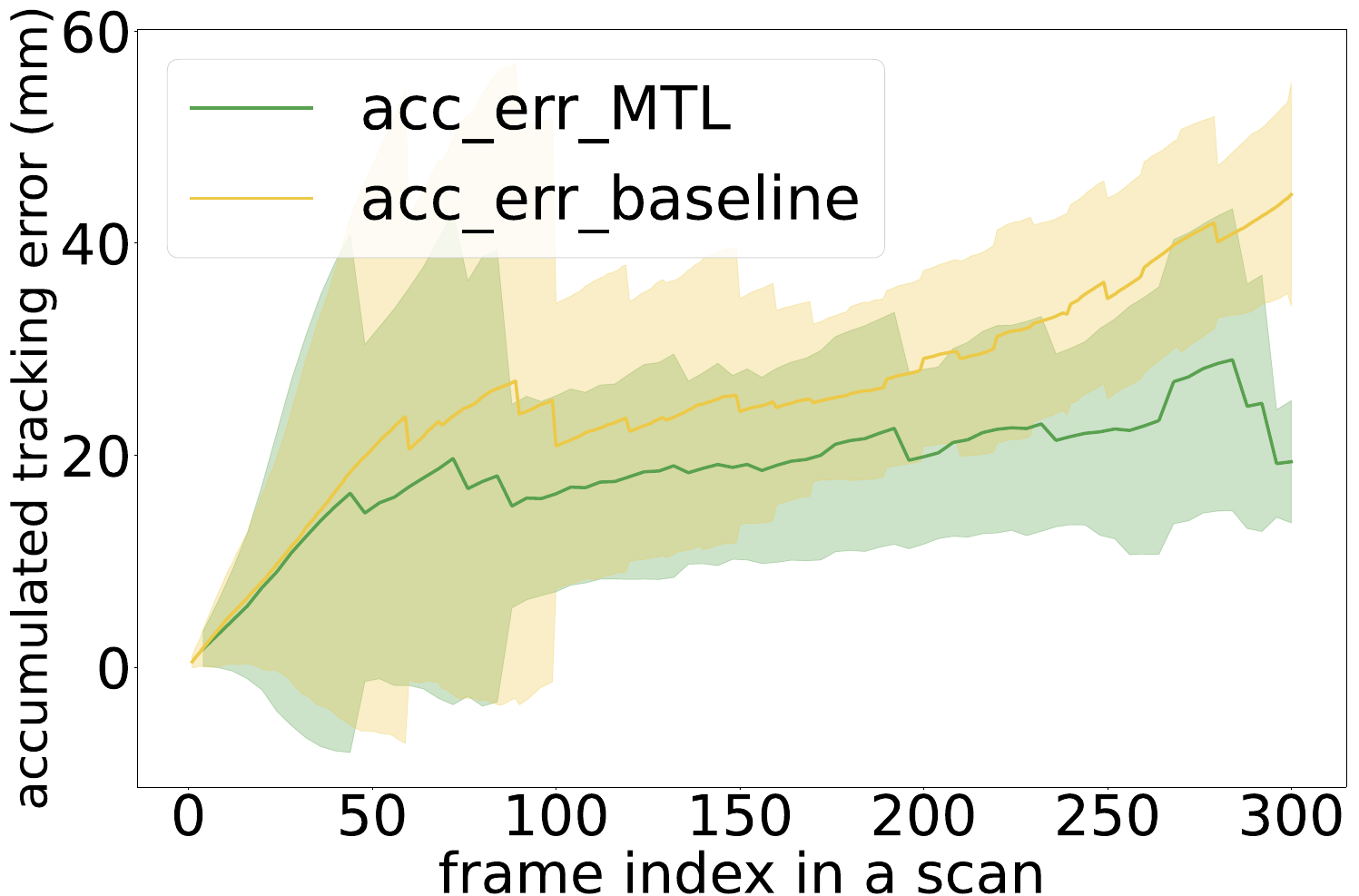}
  
  \caption{$\epsilon_{acc.}$}
  
  \label{fig:Accumulated_err}
\end{subfigure}
\caption{Comparison of $\epsilon_{frame}$ and $\epsilon_{acc.}$ between baseline and proposed method.}

\label{fig:frame_prediction_accuracy}
\end{figure}

In conclusion, the proposed trackerless freehand US improved baseline performance, by utilising sequence modelling and multi-tasking as hyperparameters, supported by a set of extensive experiments. The published code and data should also be valuable for furthering research in this area.

\section*{Compliance with ethical standards}
This study was performed in line with the principles of the Declaration of Helsinki. Approval was granted by the Ethics Committee of local institution.

\section*{Acknowledgement}
This work was supported by the EPSRC [EP/T029404/1], a Royal Academy of Engineering / Medtronic Research Chair [RCSRF1819\textbackslash7\textbackslash734] (TV), and Wellcome/EPSRC Centre for Interventional and Surgical Sciences [203145Z/16/Z]. Qi Li was supported by the University College London Overseas and Graduate Research Scholarships.

\bibliographystyle{splncs04}
\bibliography{samplepaper}

\begin{thebibliography}{10}
\providecommand{\url}[1]{\texttt{#1}}
\providecommand{\urlprefix}{URL }
\providecommand{\doi}[1]{https://doi.org/#1}

\bibitem{chen1997determination}
Chen, J., Fowlkes, J., et~al.: Determination of scan-plane motion using speckle
  decorrelation: Theoretical considerations and initial test. International
  Journal of Imaging Systems and Technology  \textbf{8}(1),  38--44 (1997)

\bibitem{guo2020sensorless}
Guo, H., Xu, S., et~al.: Sensorless freehand 3d ultrasound reconstruction via
  deep contextual learning. In: MICCAI. pp. 463--472. Springer (2020)

\bibitem{hassenpflug2005speckle}
Hassenpflug, P., Prager, R., et~al.: Speckle classification for sensorless
  freehand 3-d ultrasound. Ultrasound in medicine \& biology  \textbf{31}(11),
  1499--1508 (2005)

\bibitem{hochreiter1997long}
Hochreiter, S., Schmidhuber, J.: Long short-term memory. Neural computation
  \textbf{9}(8),  1735--1780 (1997)

\bibitem{hu2017freehand}
Hu, Y., Gibson, E., et~al.: Freehand ultrasound image simulation with
  spatially-conditioned generative adversarial networks. In: Molecular imaging,
  reconstruction and analysis of moving body organs, and stroke imaging and
  treatment, pp. 105--115. Springer (2017)

\bibitem{hu2016development}
Hu, Y., Kasivisvanathan, V., et~al.: Development and phantom validation of a
  3-d-ultrasound-guided system for targeting mri-visible lesions during
  transrectal prostate biopsy. IEEE Transactions on Biomedical Engineering
  \textbf{64}(4),  946--958 (2016)

\bibitem{jiang2021motion}
Jiang, Z., Wang, H., et~al.: Motion-aware robotic 3d ultrasound. In: ICRA. pp.
  12494--12500. IEEE (2021)

\bibitem{lasso2014plus}
Lasso, A., Heffter, T., et~al.: Plus: open-source toolkit for ultrasound-guided
  intervention systems. IEEE transactions on biomedical engineering
  \textbf{61}(10),  2527--2537 (2014)

\bibitem{luo2021self}
Luo, M., Yang, X., et~al.: Self context and shape prior for sensorless freehand
  3d ultrasound reconstruction. In: MICCAI. pp. 201--210. Springer (2021)

\bibitem{luo2022deep}
Luo, M., Yang, X., et~al.: Deep motion network for freehand 3d ultrasound
  reconstruction. In: MICCAI. pp. 290--299. Springer (2022)

\bibitem{mikaeili2022trajectory}
Mikaeili, M., Bilge, H.: Trajectory estimation of ultrasound images based on
  convolutional neural network. Biomedical Signal Processing and Control
  \textbf{78},  103965 (2022)

\bibitem{miura2020localizing}
Miura, K., Ito, K., et~al.: Localizing 2d ultrasound probe from ultrasound
  image sequences using deep learning for volume reconstruction. In: ASMUS, pp.
  97--105. Springer (2020)

\bibitem{miura2021pose}
Miura, K., Ito, K., et~al.: Pose estimation of 2d ultrasound probe from
  ultrasound image sequences using cnn and rnn. In: ASMUS. pp. 96--105.
  Springer (2021)

\bibitem{miura2021probe}
Miura, K., Ito, K., et~al.: Probe localization from ultrasound image sequences
  using deep learning for volume reconstruction. In: International Forum on
  Medical Imaging in Asia. vol. 11792, pp. 133--138. SPIE (2021)

\bibitem{mozaffari2017freehand}
Mozaffari, M., Lee, W.: Freehand 3-d ultrasound imaging: a systematic review.
  Ultrasound in medicine \& biology  \textbf{43}(10),  2099--2124 (2017)

\bibitem{ning2022spatial}
Ning, G., Liang, H., et~al.: Spatial position estimation method for 3d
  ultrasound reconstruction based on hybrid transfomers. In: ISBI. pp.~1--5.
  IEEE (2022)

\bibitem{prevost2017deep}
Prevost, R., Salehi, M., et~al.: Deep learning for sensorless 3d freehand
  ultrasound imaging. In: MICCAI. pp. 628--636. Springer (2017)

\bibitem{prevost20183d}
Prevost, R., Salehi, M., et~al.: 3d freehand ultrasound without external
  tracking using deep learning. Medical image analysis  \textbf{48},  187--202
  (2018)

\bibitem{takase2019character}
Takase, S., Suzuki, J., Nagata, M.: Character n-gram embeddings to improve rnn
  language models. In: AAAI. vol.~33, pp. 5074--5082 (2019)

\bibitem{tan2019efficientnet}
Tan, M., Le, Q.: Efficientnet: Rethinking model scaling for convolutional
  neural networks. In: International conference on machine learning. pp.
  6105--6114. PMLR (2019)

\bibitem{xie2021image}
Xie, Y., Liao, H., et~al.: Image-based 3d ultrasound reconstruction with
  optical flow via pyramid warping network. In: EMBC. pp. 3539--3542. IEEE
  (2021)

\end{thebibliography}
%
%

%\bibliographystyle{IEEEbib}
%\bibliography{strings,refs}
%\end{thebibliography}
\end{document}